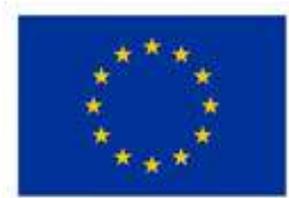

This work was carried out in whole or in part within the framework of the NOMATEN Centre of Excellence, supported from the European Union Horizon 2020 research and innovation program (Grant Agreement No. 857470) and from the European Regional Development Fund via the Foundation for Polish Science International Research Agenda PLUS program (Grant No. MAB PLUS/2018/8), and the Ministry of Science and Higher Education's initiative "Support for the Activities of Centers of Excellence Established in Poland under the Horizon 2020 Program" (agreement no. MEiN/2023/DIR/3795).

The version of record of this article, first published in Journal of Alloys and Compounds, Volume 1035, 5 July 2025, 181580, is available online at Publisher's website: https://doi.org/10.1016/j.jallcom.2025.181580

# Exploring the impact of Ti/Al on L1$_2$ nanoprecipitation and deformation behavior in CoNiFeAlTi multi-principal element alloys through atomistic simulations

Amin Esfandiarpour[1*], Anshul D. S. Parmar[1], Silvia Bonfanti[1], Pawel Sobkowicz[1], Byeong-Joo Lee[2*] and Mikko Alava[1,3]

[1] NOMATEN Centre of Excellence, National Centre for Nuclear Research, Otwock, Andrzeja Soltana 7, 05-400, Poland

[2] Department of Materials Science and Engineering, Pohang University of Science and Technology, Pohang 37673, South Korea

[3] Department of Applied Physics, Aalto University, P.O. Box 11000, 00076 Aalto, Espoo, Finland

\* Corresponding authors amin.esfandiarpour@ncbj.gov.pl (Amin Esfandiarpour)
calphad@postech.ac.kr (Byeong-Joo Lee)

**Abstract**

Recent experimental studies on CoNi-based multi-principal element alloys (MPEAs) have demonstrated high strength and ductility, attributed to the formation of stable L1$_2$ nanoscale precipitates. However, the fundamental mechanisms behind such impressive properties in these complex alloys are not well understood. In this work, we investigate the effects of Ti and Al concentrations on the formation of L1$_2$ precipitates in (CoNiFe)$_{84}$(Al$_8$Ti$_8$), (CoNiFe)$_{86}$(Al$_7$Ti$_7$), (CoNiFe)$_{88}$(Al$_6$Ti$_6$), and (CoNiFe)$_{94}$(Al$_4$Ti$_2$) MPEAs using hybrid molecular dynamics/Monte Carlo (MD/MC) simulations and a developed MEAM interatomic potential for the CoNiFeTiAl system. Additionally, we study the effect of L1$_2$ precipitation on the mechanical properties and stacking fault energy of these MPEAs using MD. Our hybrid MD/MC simulations show that (CoNiFe)$_{86}$(Al$_7$Ti$_7$) alloy exhibits the highest amount of L1$_2$ nanoprecipitates. We find that L1$_2$ precipitation increases the stacking fault energy, with higher Al and Ti contents leading to greater increases. Tensile simulations reveal that L1$_2$ precipitates enhance yield strength, with alloys exhibiting higher precipitation showing increased flow stress. We also investigate dislocation-nanoprecipitate interactions with different precipitate sizes in (CoNiFe)$_{86}$(Al$_7$Ti$_7$) alloy. Larger nanoprecipitate sizes result in stronger dislocation pinning. Dislocation-precipitate interactions indicate that dislocations predominantly shear through 4-8 nm precipitates instead of looping around them (Orowan mechanism), which enhances strength while maintaining good ductility. Although the lattice mismatch between the L1$_2$ nanoprecipitate and the matrix is low (0.139%), the significant difference in stacking fault energy between the L1$_2$ nanoprecipitate and the matrix results in stronger dislocation pinning. This fundamental understanding can guide the compositional design of MPEAs with tailored properties by controlling nanoscale precipitation.

**Keywords**: L1$_2$ nanoprecipitates, Hybrid MD/MC, SFE mismatch, strength-ductility trade-off, multi-principal element alloys

## 1. Introduction

The development of multi-principal element alloys (MPEAs), also known as high- or medium-entropy alloys (HEAs/MEAs), presents significant opportunities for materials innovation. These alloys exhibit superior physical and mechanical properties compared to conventional alloys, making them promising candidates for future structural applications [1]. However, designing MPEAs with desirable mechanical properties faces considerable challenges due to their vast composition space [2,3]. A primary challenge is the trade-off between strength and ductility in HEAs/MEAs, which complicates the design of novel MPEAs [4]. For example, HEAs/MEAs with a body-centered cubic (BCC) structure exhibit high strength at elevated temperatures but low ductility at room temperature [5]. Conversely, face-centered cubic (FCC) H/MEAs demonstrate good ductility but lower strength at high temperatures [6].

Numerous strategies for designing strong MPEAs with good ductility have been presented. For instance, some dual-phase FCC-BCC MPEAs effectively balance both ductility and strength [7–10]. Due to the high ductility of FCC-MPEAs, one strategy to enhance strong-ductile MPEAs is to increase the strength of FCC-MPEAs through solid solution strengthening. Following this strategy, FCC-MEAs with three constituent elements exhibit higher lattice distortion and strength compared to quaternary and quinary MPEAs [11–16]. Complementary to such designs, short-range order (SRO) strengthening can also be employed to increase their strength [17–21]. Additionally, nanoprecipitates (NPs) can further improve the mechanical properties of these materials [22–24]. The size and volume fraction of precipitates significantly affect both ductility and strength, with nano-sized precipitates in MPEAs overcoming the strength-ductility trade-off [22,25–27]. Specifically, $L1_2$ NPs can form in some FCC-MPEAs by adding Al and Ti [25,28–36], resulting in alloys with high strength and good ductility. Stacking fault energy (SFE) also plays a crucial role in the mechanical properties of these alloys [25,29,37]. Despite numerous recent experimental studies, a fundamental understanding of the formation of such precipitates and their effect on the mechanical properties of these complex alloys remains lacking.

In this study, we construct a MEAM interatomic potential for CoNiFeTiAl to investigate how varying atomic percentages of Al and Ti influence the formation of the $L1_2$ phase in several CoNiFeAlTi alloys, using a hybrid molecular dynamics/Monte Carlo (MD/MC) method. Hybrid MD/MC is a widely used method to investigate the ordering phase after annealing in High/Medium Entropy Alloys (H/MEAs) [13,17,18,38–41]. For instance, SRO in CoNiCr, as observed in experimental studies [42], was detected using this method [17,18,39]. Furthermore, the formation of the $L1_2$ phase ordering was studied in $Fe_xNi_{1-x}$ alloys with hybrid MD/MC [41], yielding results consistent with experimental findings [43]. We performed tensile tests to investigate the effect of NPs on yield strength and flow stress. Using molecular statics, we examined the dependency of SFE on Al and Ti concentrations. Additionally, we investigated the dependency of SFE on the concentration of $L1_2$. Finally, using shear control loading, we examined the effect of $L1_2$ NPs on dislocation pinning.

## 2. Methods

All molecular dynamics (MD) simulations conducted in this study utilized the large-scale atomic/molecular massively parallel simulator (LAMMPS) [44]. In this study, all simulations employed the second nearest-neighbor modified embedded atom method (2NN MEAM) interatomic potential. Modeling the CoNiFeAlTi quinary system required potential parameter sets for the constituent unary, binary, and ternary systems. Published potentials were used for the pure elements (Co [45], Ni [45], Fe [46], Al [47], Ti [48]) as well as the binary pairs (Co-Fe [49], Co-Ni [50], Co-Al [50], Co-Ti [51], Ni-Fe [52], Ni-Al [53], Ni-Ti [53], Fe-Al [54], Fe-Ti [55], Al-Ti [53]). Ternary potentials for Co-Ni-Fe [49], Co-Ni-Al [50], Ni-Fe-Al [56], and Ni-Al-Ti [53] systems were also available from prior works. However, potentials for the remaining ternary combinations had not been previously developed. We derived the missing potential parameters using an approach analogous to Choi et al [49]. Modeling each ternary system involved determining six screening parameters: three Cmin(i-j-k) and three Cmax(i-j-k) values. These parameters describe the screening effect of an atom j on the interactions between neighboring atoms i and k. The screening parameters were assigned values based on an averaging concept. We listed all these parameters for the ternary systems in Table S1. With all required potentials determined, the complete quinary interatomic potential was obtained. The parameter values for this potential in LAMMPS format are provided as supplementary information. In addition, to validate this potential, the elastic constant and stacking fault energy of the (Ni,Co,Fe)$_3$(Ti,Al,Fe) L1$_2$ phase that formed in (CoNiFe)$_{86}$(Al$_7$Ti$_7$) alloy were compared with results obtained using density functional theory (DFT) method [57] (see Table S2). The results show good agreement between the MEAM and DFT data.

To investigate the effect of Al and Ti on L1$_2$ phase formation in CoNiFeAlTi alloys, we conducted hybrid MD/MC atomistic simulations at temperature 300 K and 1100 K. The simulations followed the hybrid MD/MC approach described in Refs. [4,5]. Initially, Co, Ni, Fe, Al and Ti atoms were randomly distributed in a FCC simulation box with dimensions of 110.49 Å along <100>, 242.36 Å along <010>, and 114.05 Å along <001>, which was then exposed to energy minimization via the conjugate gradient method. The system was equilibrated at the target temperature under NPT ensemble conditions. The subsequent MC swap methodology involved randomly selecting two atoms of different elemental types and attempting to swap their identities while conserving the total kinetic energy. The acceptance criterion for each attempted swap move followed the Metropolis probability: $u \leq \min(1, \exp[-\beta \Delta U])$, where $u$ is a uniform random number between 0 and 1, $\Delta U$ is the potential energy change from the swap, $\beta = 1/k_B$, and $k_B$ is the Boltzmann constant. Following a sequence of 200 swap attempts, the system was relaxed for 50 molecular dynamics steps under NPT conditions. This cycle of swap attempts and relaxations continued iteratively until the system's potential energy satisfactorily converged. A total of 14 million swap attempts were made. To analyze the ordering type, we calculated the total percentage of L1$_2$ atoms using the polyhedral template matching method [58]. We considered the Al and Ti atoms as the first group and the Co, Ni, and Fe atoms as the second group. The total L1$_2$ atoms were then extracted based on this grouping.

To see the effect of L1$_2$ on the tensile properties and SFE, we created FCC samples with the orientation $X = [1\bar{1}2]$, $Y = [\bar{1}11]$, and $Z = [\bar{1}\bar{1}0]$, containing 55296 atoms with approximate dimensions $69.6 \times 147.6 \times 60.3$ Å$^3$.

We conducted uniaxial tensile tests using MD both before and after hybrid MD/MC simulations in the y direction. The NPT ensemble was employed during the deformation process, with the pressure in the direction perpendicular to the stretching maintained at zero. Periodic boundary conditions (PBC) were applied in all directions. The tensile tests were performed at 300 K, with strain rates set at $\dot{\varepsilon} = 2 \times 10^8 \, 1/s$ and $\dot{\varepsilon} = 2 \times 10^9 \, 1/s$.

To calculate the stacking fault energy, we obtained the generalized planar fault energy (GSFE) curves by incrementally displacing the top half of the simulation box by 0.01 Å along the (111) plane in the [11$\bar{2}$] direction until a stacking fault formed. After each increment, the configuration underwent constrained energy minimization, allowing the system to relax only along the y direction. For visualization, we used OVITO software [59].

## 3. Results and discussion

### 3-1. Effect of Ti and Al in the formation of L1$_2$

In this study, we employed hybrid MD/MC to investigate the formation of L1$_2$ NPs in four MPEAs: (CoNiFe)$_{84}$(Al$_8$Ti$_8$), (CoNiFe)$_{86}$(Al$_7$Ti$_7$), (CoNiFe)$_{88}$(Al$_6$Ti$_6$), and (CoNiFe)$_{94}$(Al$_4$Ti$_2$).

Fig. 1 presents these results. Fig. 1a shows the distribution of atoms after 14 milion swaps at 300 K, where a Ni-rich L1$_2$ phase formed in all four MPEAs. A Ni-rich L1$_2$ phase was also observed in experimental studies for (CoNiFe)$_{86}$(Al$_7$Ti$_7$) MPEA [60]. Figs. 1b and 1d show the results at 300 K and 1000 K, respectively. (CoNiFe)$_{86}$(Al$_7$Ti$_7$) alloy contains the highest amount of L1$_2$, while (CoNiFe)$_{94}$(Al$_4$Ti$_2$) alloy has the lowest. The L1$_2$ atomic percentage decreases with increasing annealing temperature. The formation of the L1$_2$ nanoprecipitate (NP) phase in these alloys is evidenced by the system reaching its lowest potential energy (see Figs. 1c and 1e). The spatial distribution of the L1$_2$ phase across different alloy compositions is shown in Fig. S1.

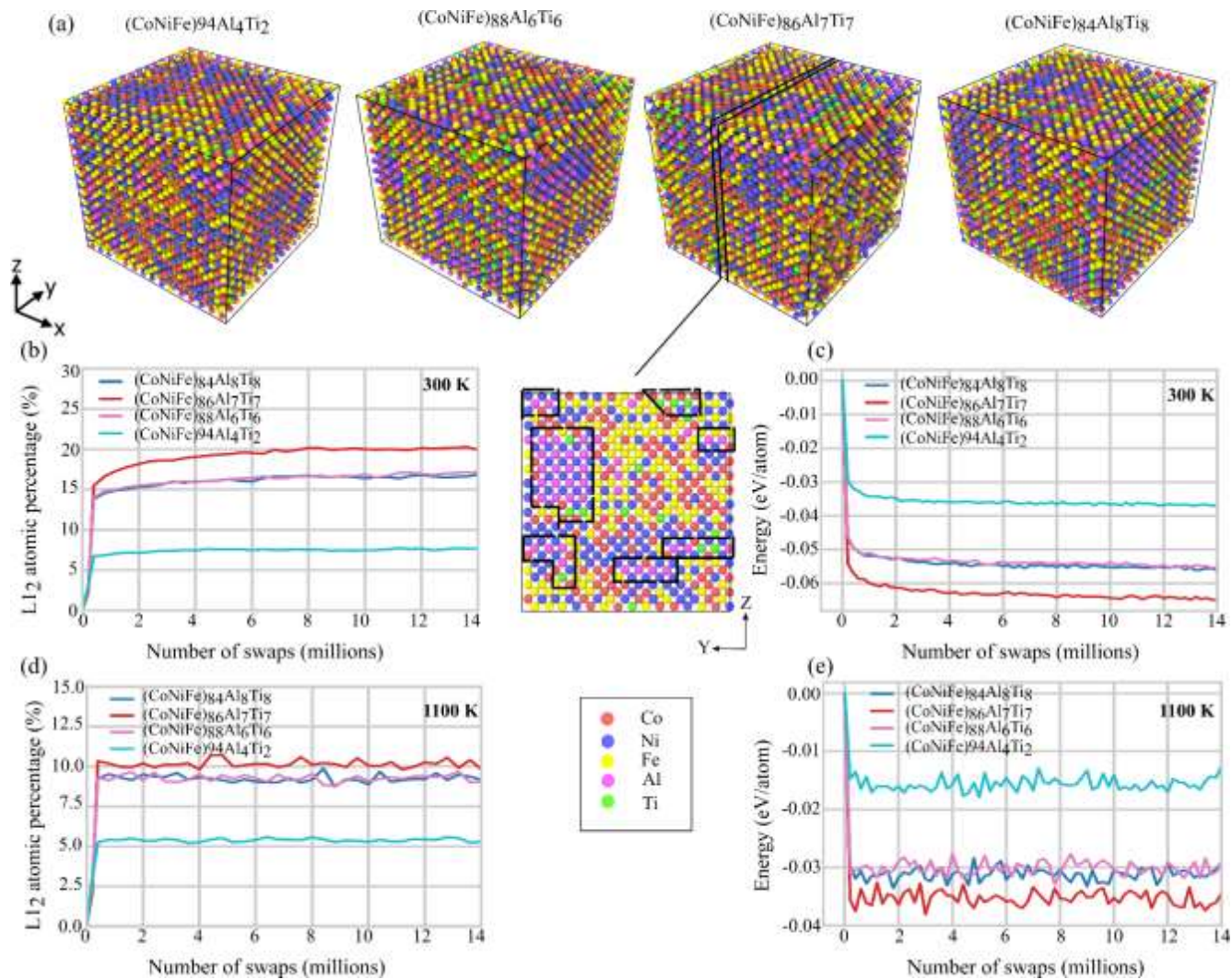

Fig. 1: Hybrid MD/MC simulation outcomes for four CoNiFeAlTi alloys, based on 14 million swap attempts: (a) Final atomic distribution at 300 K. (b-c) Variation in L1$_2$ atomic percentage and associated potential energies with the number of swaps at 300 K. (d-e) Variation in L1$_2$ atomic percentage and associated potential energies with the number of swaps at 1000 K

Although our simulations did not explicitly calculate the isolated formation energy of the L1$_2$ phase itself, the significantly lower equilibrium potential energy observed at (CoNiFe)$_{86}$(Al$_7$Ti$_7$), compared to other compositions tested, strongly suggests that the formation of L1$_2$ at this composition is energetically favorable. Thus, the highest observed L1$_2$ fraction at (CoNiFe)$_{86}$(Al$_7$Ti$_7$) is indeed correlated with the alloy's enhanced energetic preference for ordering. Additionally, we have conducted a detailed analysis of the SRO using the Warren–Cowley parameters (WCP). The results reveal a strong ordering tendency among Ni–Al and Co–Ti atomic pairs in the first-nearest-neighbor shell. Among the investigated compositions, the (CoNiFe)$_{86}$(Al$_7$Ti$_7$) alloy exhibits the most pronounced Ni–Al ordering, while the (CoNiFe)$_{94}$(Al$_4$Ti$_2$) composition shows the weakest. These findings are consistent with the emergence of Ni-rich L1$_2$-type local structures, in agreement with our calculations of the L1$_2$ phase. The computational details of the WCP analysis and a heatmap visualizing the results are included in the Supplementary Information (Fig. S2).

The variation in L1$_2$ phase volume fraction observed in our hybrid MD/MC simulations may reflect similar tendencies reported in earlier studies of CoNi-based MPEAs. For instance, studies on equimolar Co-Ni-based MPEAs with added Al and Ti report that the L1$_2$ volume fraction typically does not exceed 30% [60–62]. Additionally, adjustments in the Al/Ti ratio have been shown to influence phase distribution [62,63]. Specifically, Ding et al. demonstrated that a higher Al/Ti ratio could reduce the L1$_2$ phase volume fraction [62], which is consistent with our observation of a lower L1$_2$ fraction in (CoNiFe)$_{94}$(Al$_4$Ti$_2$) relative to (CoNiFe)$_{86}$(Al$_7$Ti$_7$). Another study indicates that increasing Ti content while keeping Al constant can raise the L1$_2$ phase volume fraction [63]. Although both Al and Ti contents change between (CoNiFe)$_{86}$(Al$_7$Ti$_7$) and (CoNiFe)$_{88}$(Al$_6$Ti$_6$) the relatively higher Ti content in the (CoNiFe)$_{86}$(Al$_7$Ti$_7$) alloy may partially contribute to its increased L1$_2$ phase fraction, in line with the trend reported in Ref. [63]. Furthermore, (CoNiFe)$_{86}$(Al$_7$Ti$_7$) has a higher L1$_2$ fraction than (CoNiFe)$_{84}$(Al$_8$Ti$_8$), which could be due to the effect of high Al concentrations in FCC HEAs. Excess Al in these alloys often promotes the formation of the B2 phase rather than the L1$_2$ phase [64]. These trends highlight that (CoNiFe)$_{86}$(Al$_7$Ti$_7$), with its balanced Al and Ti content, achieves the optimal conditions for a high L1$_2$ phase volume fraction, while alloys with higher Al or Al/Ti ratios (such as (CoNiFe)$_{94}$(Al$_4$Ti$_2$) and (CoNiFe)$_{84}$(Al$_8$Ti$_8$)) show reduced L1$_2$ fractions, consistent with our findings.

### 3-2 Tensile properties

To study the effect of NPs on tensile properties, we conducted uniaxial tensile tests using MD on (CoNiFe)$_{86}$(Al$_7$Ti$_7$) and (CoNiFe)$_{94}$(Al$_4$Ti$_2$) alloys before and after hybrid MD/MC simulations. These two alloys were selected because they represent the highest and lowest L1$_2$ phase percentages among the four different alloys studied. For ease of identification, we refer to the ordered and random versions of (CoNiFe)$_{84}$(Al$_8$Ti$_8$), (CoNiFe)$_{86}$(Al$_7$Ti$_7$), (CoNiFe)$_{88}$(Al$_6$Ti$_6$), and (CoNiFe)$_{94}$(Al$_4$Ti$_2$) MPEAs as Al8Ti8-O and Al8Ti8-R, Al7Ti7-O and Al7Ti7-R, Al6Ti6-O and Al6Ti6-R, and Al4Ti2-O and Al4Ti2-R, respectively. To ensure robust statistics, we prepared 8 different samples (using different random seeds for initial distributions) for each alloy before and after hybrid MD/MC simulations. After hybrid MD/MC, we obtained 8 samples with nearly identical potential energies and L1$_2$ phase percentages but different local positions of L1$_2$ NPs.

Tensile tests were performed at 300 K at two different strain rates: $\dot{\varepsilon} = 2 \times 10^8 1/s$ and $\dot{\varepsilon} = 2 \times 10^9 1/s$. Fig. 2a shows the direction of loading. Fig. 2 illustrates the results for Al7Ti7 at these two strain rates. Figs. 2b and 2c present the stress-strain curves for Al7Ti7-O and Al7Ti7-R (mean values for 8 samples) at 300 K under the two strain rates. Figs. 2d and 2e depict the evolution of dislocations during loading.

The strengthening mechanisms in Co-Ni-based MPEAs arise from NP strengthening and solid solution strengthening, both of which rely on dislocation interactions with precipitates or solute atoms. In tensile testing of single-crystal pristine samples without initial dislocations, the yield point reflects a tendency for initial dislocation formation, while the flow stress represents the subsequent growth and interaction of these dislocations with NPs.

The stress-strain curves (Figs. 2b and 2c) demonstrate that the formation of an ordered phase containing L1$_2$ NPs increases yield strength and flow stress. At the lower strain rate ($\dot{\varepsilon} = 2 \times 10^8 \, 1/s$), the ordered phase (Al7Ti7-O) exhibits a neck, whereas the random distribution of atoms (Al7Ti7-R) shows a sudden drop in stress. This increase in yield strength and flow stress due to the formation of ordered phases in Co-Ni-based MPEAs has been previously reported [18,65].

To correlate this tensile behavior with dislocation dynamics, we analyzed the total dislocation length during tension (Figs. 2d and 2e). No clear, inferable difference was observed between the total dislocation lengths of Al7Ti7-O and Al7Ti7-R alloys during loading. Fig. 2 also shows that increasing strain rate results in higher yield strength and total dislocation length for both alloys. We further analyzed the amounts of Shockley dislocations and Stair-rod dislocations during tension (Fig. 3). This figure shows that the density of Stair-rod dislocations is significantly higher in Al7Ti7-O, whereas the density of Shockley dislocations is similar in both Al7Ti7-O and Al7Ti7-R. The formation of Stair-rod dislocations, which are sessile, can contribute to the increased strength in MPEAs [18,66–68]. The majority of dislocations formed in FCC alloys are glissile partial dislocations. Stair-rod dislocations typically form when partial dislocations on intersecting {111} planes react as follows [69]:

$$\frac{a}{6}[112] + \frac{a}{6}[1\bar{1}2] \rightarrow \frac{a}{6}[001] \text{ (stair-rod dislocation)} \quad (1)$$

L1$_2$ NPs may facilitate stair-rod dislocation formation by promoting the intersection of partial dislocations in their vicinity. This may be due to the high stacking fault energy (SFE) of the NPs (see Section 3.3), which, while locally suppressing the dissociation of partial dislocations, increases dislocation activity and intersection events in the surrounding matrix. This enhanced activity can, in turn, promote stair-rod formation. This explains the higher flow stress of the ordered Al7Ti7 alloy compared to the random alloy. Additionally, dislocation nodes developed by dislocation entanglement were also observed (Fig. 3c).

Fig. 4 compares the tensile properties of Al7Ti7-O and Al4Ti2-O alloys at two different strain rates. This figure further confirms the higher flow stress of Al7Ti7-O, which has a higher amount of L1$_2$ NPs. To understand the mechanical response differences between the Al4Ti2-O and Al7Ti7-O alloys, we compared their stress-strain curves before NP formation, using the Al4Ti2-R and Al7Ti7-R alloys with a random atomic distribution inside the crystal cell (see Fig. S4). In this random configuration, both the flow and yield stresses of the Al4Ti2-R alloy are higher than those of the Al7Ti7-R alloy. However, after NP formation, the Al7Ti7-O alloy exhibits a higher flow stress than Al4Ti2-O, even though Al4Ti2 maintains a higher yield stress (Fig. 4). To explain why Al4Ti2 alloys have a higher yield stress than Al7Ti7 alloys, we examined the tendency for dislocation formation by calculating the energy factors K, based on each alloy's elastic constants [70,71]:

$$K_{screw} = \left[\frac{C_{44}(C_{11}-C_{12})}{2}\right]^{\frac{1}{2}} \quad (2)$$

$$K_{edge} = (C_{11} + C_{12})[C_{44}(C_{11} - C_{12})/C_{11}(C_{11} + C_{12} + 2C_{44})]^{\frac{1}{2}} \quad (3)$$

For Al4Ti2-R alloy ($C_{11} = 253.38 \, GPa, C_{12} = 161.53 \, GPa, C_{44} = 97.74 \, GPa$) we obtained $K_{screw} = 66.99 \, GPa$ and $K_{edge} = 99.96 \, GPa$. For Al7Ti7-R alloy ($C_{11} = 238.79 \, GPa, C_{12} = 159.91 \, GPa, C_{44} = 95.79 \, GPa$) we obtained $K_{screw} = 61.46 \, GPa$ and $K_{edge} = 92.31 \, GPa$. A smaller $K$ value signals that the alloy has a higher tendency for dislocation formation. These lower energy factors for Al7Ti7-R alloy suggest a higher tendency for dislocation formation compared to Al4Ti2-R alloy, explaining the observed differences in yield stress between the alloys.

It should be noted that while the introduction of L1$_2$ NPs increases the yield strength of both Al7Ti7 and Al4Ti2 alloys, the magnitude of yield strength increase in Al7Ti7-O is not sufficient to overcome the baseline strength difference between the corresponding random alloys. In other words, although the yield strength of Al7Ti7-O benefits from L1$_2$ phase ordering, it remains lower than that of Al4Ti2-O due to its origin from a mechanically weaker matrix (Al7Ti7-R).

### 3-3 Stacking fault energy

Stacking fault energy is one of the important properties that play a key role in the mechanical behavior of FCC MPEAs [65,68,72–74]. The SFE was calculated by displacing the atomic plane by a/6[112] (a is lattice constant) in the upper half of the simulation box. In $\gamma(FCC)/\gamma'(L1_2)$ alloys, this type of SFE is also referred to as complex SFE (CSFE) [57]. The first local minimum on the GSFE curve represents the most stable configuration and is identified as the SFE (Fig. 5b). To examine the dependency of SFE on the local environment, the Y coordinate of the moving plane was varied within the range of *[-20, 20]* (see Fig. 5a). Figs. 5c-f display the stacking fault energy for the four alloys, both with random atomic distributions and with ordered L1$_2$ NPs. Table 1 provides the mean SFE values for each alloy.

As shown in Fig. 5 and Table 1, the presence of L1$_2$ NPs in these alloys leads to an increase in SFE. Previous studies have also reported an increase in SFE with the degree of SRO [39,65,75]. Furthermore, Table 1 reveals that increasing the aluminum and titanium content from Al4Ti2 to Al8Ti8 results in higher stacking fault energy. The rise in SFE with the addition of Al is well-documented in FCC systems [76,77].

Table 1: Mean value of SFE (mJ/m$^2$) across various Y coordinates of the stacking fault plane for the four alloys, both with random atomic distributions and with ordered L1$_2$ NPs.

|  | Al4Ti2 | Al6Ti6 | Al7Ti7 | Al8Ti8 |
|---|---|---|---|---|
| SFE before hybrid MD/MC | 19.54 ± 5.19 | 21.28 ± 7.13 | 25.53 ± 4.35 | 31.28 ± 8.01 |
| SFE after hybrid MD/MC | 86.07 ± 5.63 | 97.21 ± 9.11 | 111.58 ± 10.46 | 109.97 ± 6.97 |

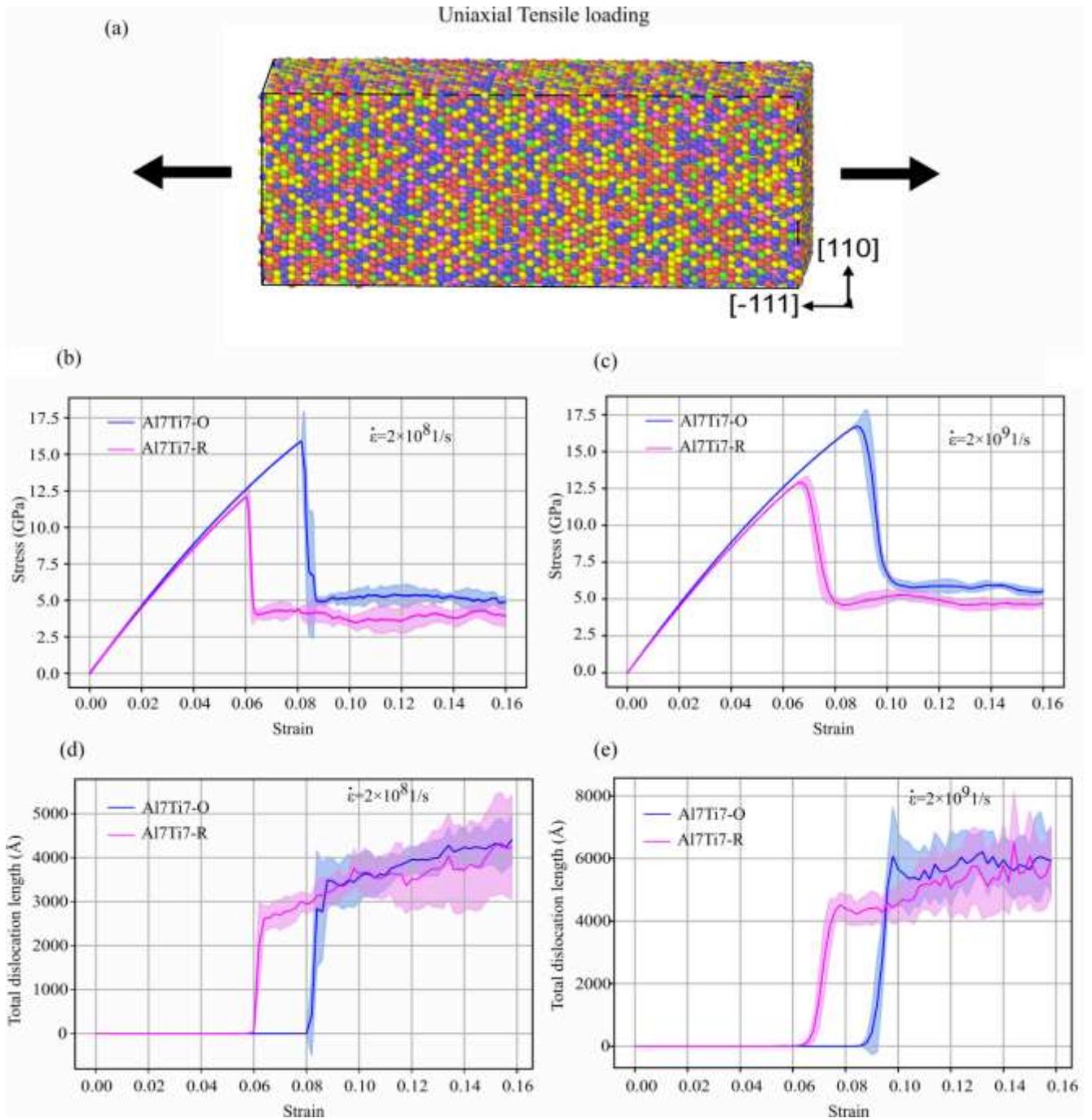

Fig. 2: Uniaxial tensile loading, stress-strain curves, and strain-dislocation length curves for $(CoNiFe)_{86}(Al_7Ti_7)$ at 300K (a) Schematic of the simulation setup. (b-c) Stress-strain curves for Al7Ti7-O and Al7Ti7-R (mean values and standard deviations) under two strain rates. (d-e) Total dislocation length vs. strain for Al7Ti7-O and Al7Ti7-R (mean values and standard deviations) under two strain rates.

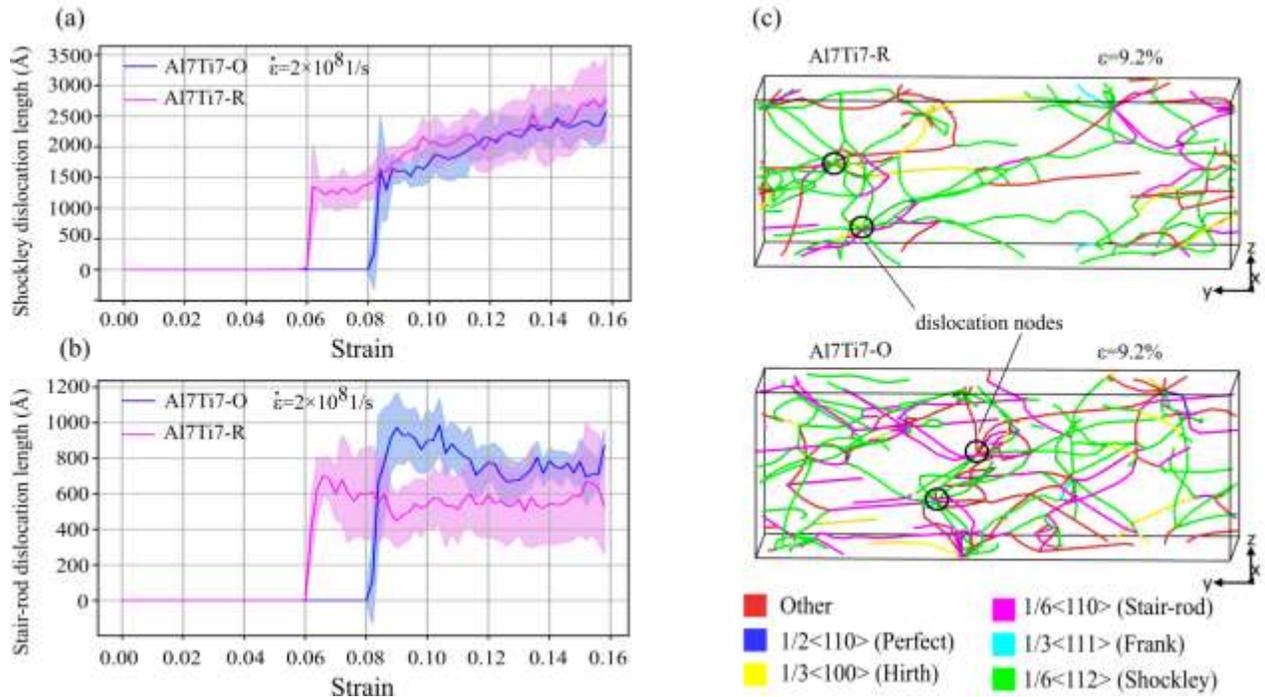

Fig. 3: (a-b) Comparison of Shockley dislocation lengths and Stair-rod dislocations during loading for Al7Ti7-O and Al7Ti7-R alloys (c) Visualization of dislocations at $\varepsilon = 9.2\%$ for Al7Ti7-O and Al7Ti7-R alloys.

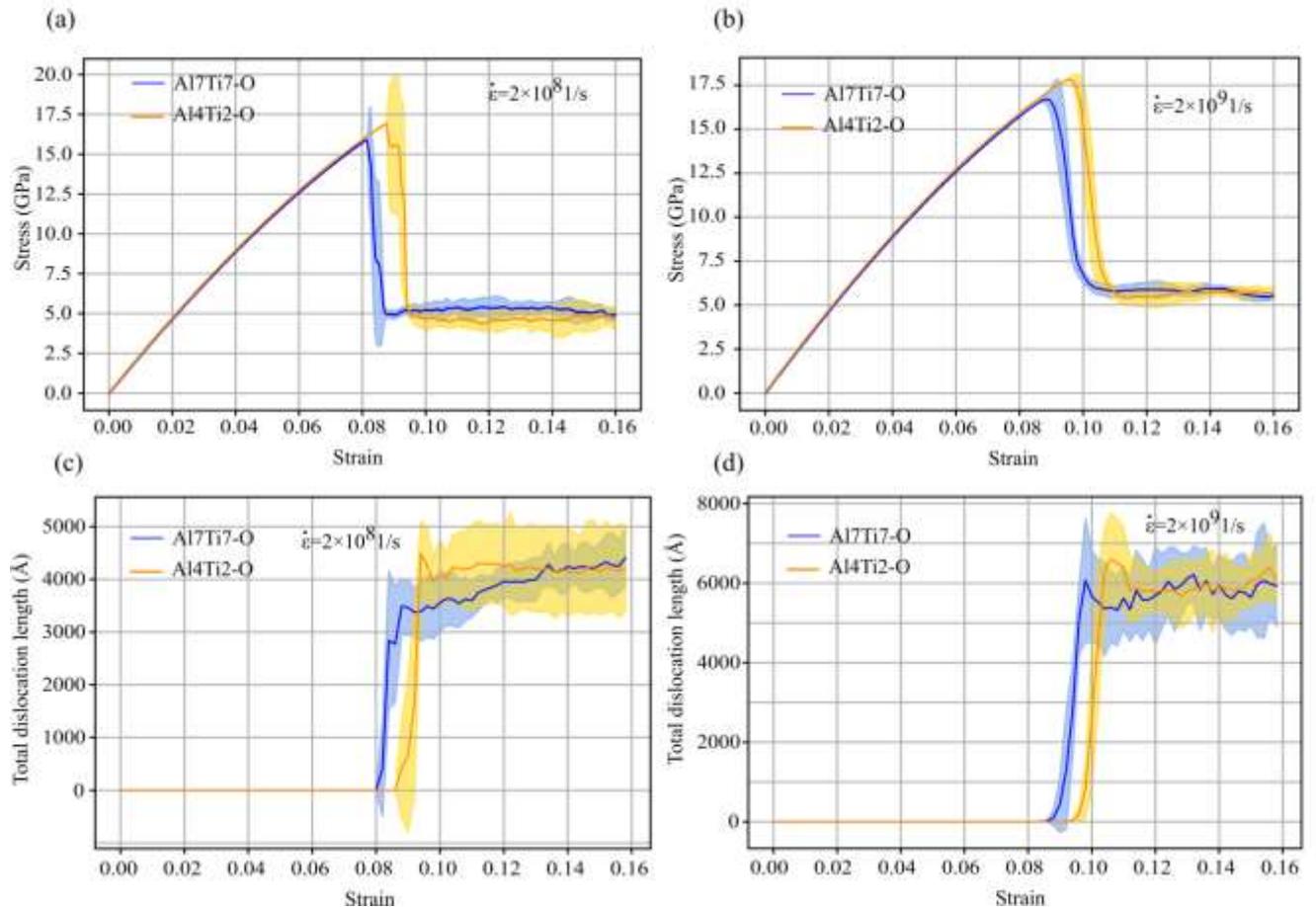

Fig. 4: (a-b) Stress-strain curves for Al7Ti7-O and Al4Ti2-O (mean values and standard deviations) under two tensile strain rates at 300 K. (d-e) Total dislocation length vs. strain for Al7Ti7-O and Al4Ti2-O (mean values and standard deviations) under two strain rates.

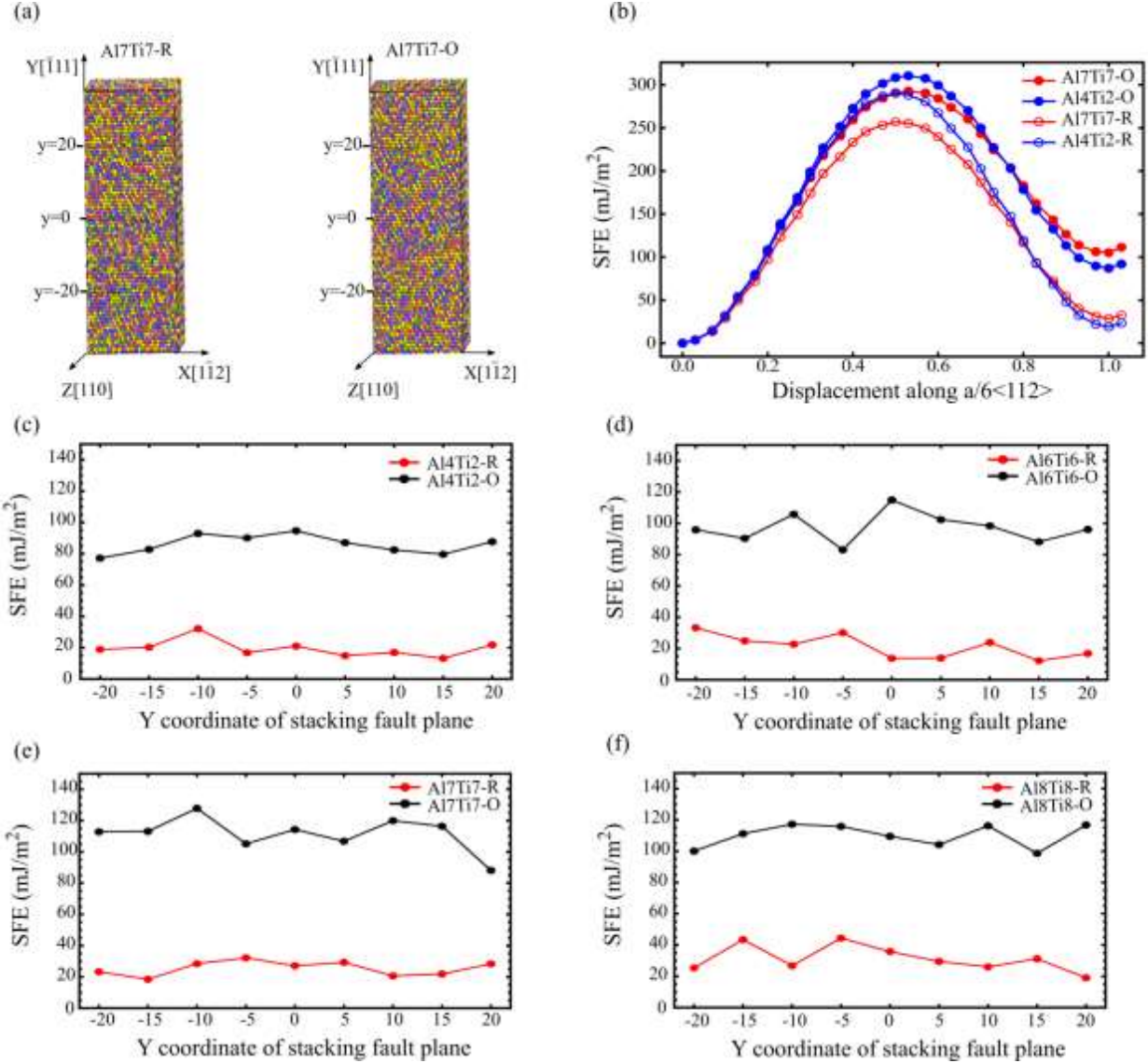

Fig. 5: (a) Schematic of the simulation box showing the Y coordinate of the moving plane within the range of *[-20, 20]* to examine the dependency of stacking fault energy (SFE) on the local environment. (b) GSFE curves for Al7Ti7-O and Al7Ti7-R, Al4Ti2-O, and Al4Ti2-R alloys for a representative Y value within the specified range. (c-f) Stacking fault energy for the four alloys, both with random atomic distributions and with ordered $L1_2$ NPs.

### 3-4 Dislocation-nanoprecipitate interaction

To investigate how $L1_2$ NPs affect dislocation pinning, an edge dislocation was introduced into the simulation box for the $(CoNiFe)_{86}Al_7Ti_7$ alloy. This composition was chosen because it contains the highest amount of $L1_2$. After relaxation, this edge dislocation dissociated into two partial dislocations, referred to as the trailing and leading partial dislocations. NPs with diameters of 4, 6, and 8 nm were positioned relative to the dislocation (see Fig. 6). The composition of the matrix and L1$_2$ NPs was extracted from Ref. [60], where the L1$_2$ NP has a (Ni,Co,Fe)$_3$(Ti,Al,Fe) structure. In our model, a spherical L1$_2$ NP was embedded within a chemically random FCC matrix. A pure Ni crystal was first constructed and oriented along $X = [1\bar{1}0]$, $Y = [11\bar{2}]$ and $Z = [111]$ directions. A spherical region was then carved out and replaced with an L1$_2$-structured Ni$_3$Al-type domain of the same diameter and lattice constant to ensure coherency. Within this sphere, 32% of Ni atoms were substituted with Co, 12% with Fe, and the remainder kept as Ni (56%). Within this sphere, 32% of Ni atoms were substituted with Co, 12% with Fe, and 56% remained as Ni. On the Al sublattice, 54% of Al atoms were replaced by Ti, 11% by Fe, and 35% remained as Al. The overall elemental composition of the L1$_2$ NP was 42% Ni, 24% Co, 11.7% Fe, 8.8% Al, and 13.5% Ti, which is close to the L1$_2$ composition reported in Ref. [60]. Outside the NP, Co, Fe, Al, and Ti were randomly substituted into the Ni matrix to achieve the target alloy composition of (CoNiFe)$_{86}$(Al$_7$Ti$_7$). Due to the low lattice misfit between the ordered L1$_2$ phase and the surrounding matrix, the interface was coherent and crystallographically aligned.

The simulation box has dimensions of 511.1 Å × 100.3 Å × 620.8 Å, oriented with $X = [1\bar{1}0]$, $Y = [11\bar{2}]$ and $Z = [111]$. To study the interaction between NPs and dislocations, shear strain-controlled loading at a rate of $8 \times 10^6\ s^{-1}$ was applied. Periodic boundary conditions were used in X and Y directions, with a fixed boundary condition in the Z direction. The Z-directional volume was divided into three regions: a central area containing typical MD mobile atoms, positioned between fixed upper and lower regions. The fixed upper and lower regions consist of several atomic layers (grey layers in Fig. 6). After relaxation, a constant strain rate was implemented through shear strain loading, achieved by applying velocity along the positive x direction to the upper region while keeping the lower region fixed. During this loading process, the microcanonical ensemble (NVE) was used in conjunction with Langevin dynamics in the central region. To minimize the potential impact of local thermal activation on dislocation behavior, Langevin dynamics maintained the system temperature at approximately 5 K.

The results are shown in Fig. 7. The results are compared with equimolar CoNiFe without any ordered phase. This figure show that increasing NP size from 4 to 8 nm in (CoNiFe)$_{86}$Al$_7$Ti$_7$ alloy led to an increase in depinning stress at 5 K, rising from 340 MPa to 450 MPa. No Orowan-looping was observed after the dislocation passed through the L1$_2$ NPs in this alloy. The only mechanism observed was dislocation cutting within this NP size range (Fig. 8). We calculated the atomic mismatch percentage between the L1$_2$ NPs and the FCC matrix using the following formula: $\delta = \frac{2(a_{L1_2} - a_{FCC})}{a_{L1_2} + a_{FCC}} \times 100$, where $a_{FCC}$ represents the lattice constant for the random composition and $a_{L1_2}$ represents the lattice constant for the site with the L1$_2$ structure. To calculate the lattice constant, we analyzed the first peak of the radial distribution function g(r) for both systems: a random FCC with (CoNiFe)$_{86}$Al$_7$Ti$_7$ composition and a system with the corresponding L1$_2$ composition after minimization at 0 K (Fig. S3). By considering the pair separation distance at the first peak as the first nearest neighbor atom distance ($a/\sqrt{2}$) in these compositions, we calculated $a_{FCC} = 3.596$ Å and $a_{L1_2} = 3.601$ Å. This results in $\delta = 0.139\%$ which is very close to the experimental lattice mismatch for (CoNiFe)$_{86}$Al$_7$Ti$_7$ MPEA(0.21%) [60]. The small positive lattice mismatch indicates the formation of coherent NPs in (CoNiFe)$_{86}$Al$_7$Ti$_7$. This small lattice mismatch alone should not be the primary reason for the increase in depinning stress by approximately 110 MPa when the size of the NPs increases from 4 to 8 nm. Therefore, we sought a more substantial physical explanation. Figs. 7b-h provide an interesting clue, showing that the distance between the two partial dislocations inside and outside the NP differs. Given that this distance is inversely proportional to the stacking fault energy [75], it suggests that the SFE in the L1$_2$ NP is higher than in the matrix. Increases in SFE and decreases in dislocation distance with increasing SRO have been reported in experimental results for VCoNi [75]. We calculated the stacking fault energy for a system with the same composition as the L1$_2$ NPs using the same method described in the previous section. Additionally, we calculated the stacking fault energy mismatch as follows: $\delta_{SFE} = \frac{2(SFE_{L1_2}-SFE_{FCC})}{SFE_{L1_2}+SFE_{FCC}}$. We calculated $SFE_{L1_2} = 337.36 \pm 13.95 \, mJ/m^2$, which is consistent with DFT values reported for the (Ni,Co,Fe)$_3$(Ti,Al,Fe) L1$_2$ structure ($354 \, mJ/m^2$) [57] . By extracting $SFE_{FCC}$ from Table 1 for the random distribution of (CoNiFe)$_{86}$Al$_7$Ti$_7$ alloy ($25.53 \pm 4.3 \, mJ/m^2$), we calculated $\delta_{SFE} = 1.178$, representing a large SFE mismatch (117.8%) between the precipitate and the matrix. The difference in stacking fault energy between the matrix and NPs induces a pinning force as dislocations pass through, despite the small lattice mismatch between the NPs and the matrix. This results are supported by other studies, which have shown that the strength of alloys can be increased by larger fluctuations in the stacking fault energy [78,79]. These results suggest that a low lattice mismatch combined with a high stacking fault energy mismatch between the NPs and the matrix can be an effective strategy for increasing strength without sacrificing ductility. We did not observe any Orowan loops after dislocations passed through L1$_2$ precipitates with sizes up to 8 nm. The relative size of the dislocation line and NPs may influence the depinning stress and potentially alter the interaction mechanism [80,81]. To ensure that the dislocation line length does not affect the observed mechanism, we performed additional simulations with a longer dislocation line (Y ≈ 32 nm) interacting with an 8 nm L1$_2$ NP. No evidence of Orowan looping was observed, even after multiple dislocation-NP interactions.  Fig. 8 shows the cutting of precipitates after multiple dislocation passes. The low lattice misfit may be responsible for the absence of Orowan loops.

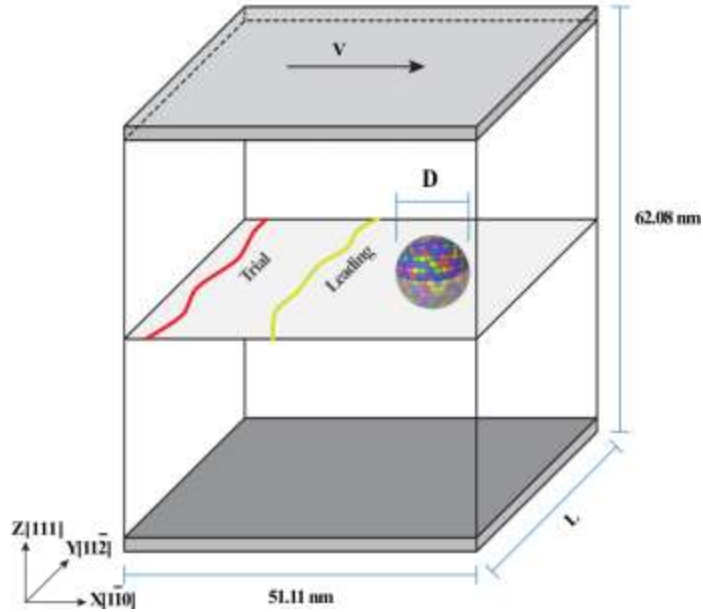

Fig. 6: Schematic of the simulation box used for performing shear strain control loading to study dislocation-nanoprecipitate interaction in (CoNiFe)$_{86}$Al$_7$Ti$_7$ alloy .

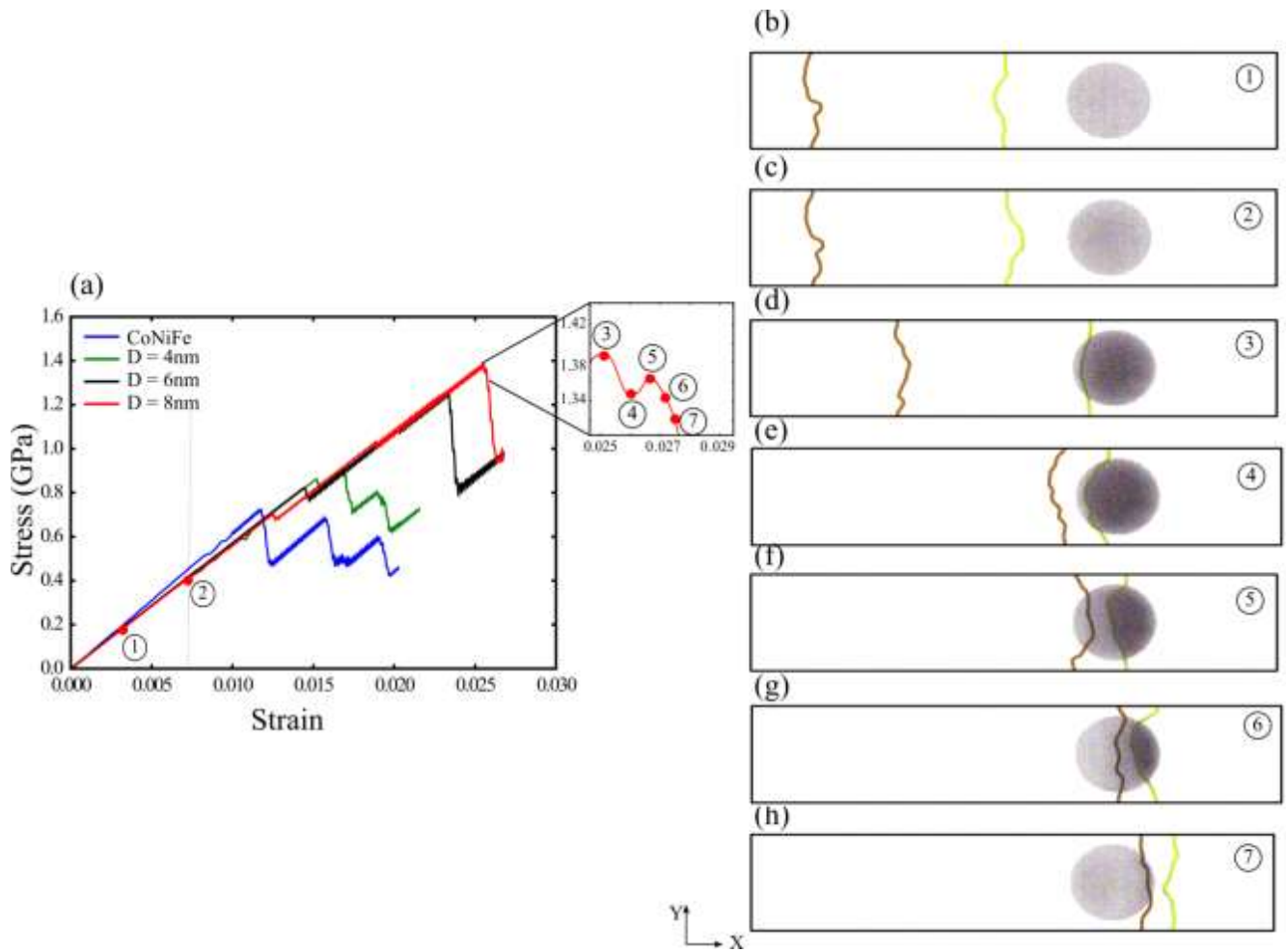

Fig. 7: (a) Stress–strain curves under shear loading for $(CoNiFe)_{86}Al_7Ti_7$ containing an edge dislocation and a spherical nanoprecipitate (NP) of varying diameters. For comparison, the shear response of equimolar CoNiFe with an edge dislocation but without NPs is also shown. (b–h) Snapshots of the simulation box illustrating the dislocation and an 8 nm NP at points 1–7 indicated on the stress–strain curve.

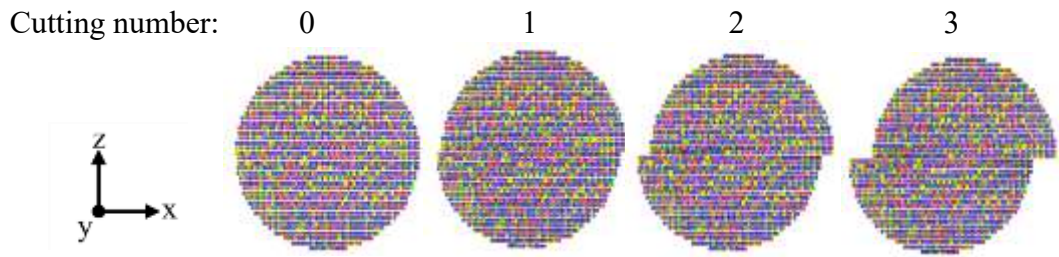

Fig. 8: Snapshot of a nanoprecipitate with a diameter of 8 nm after several passes of a dislocation (cutting numbers) through it

## 4. Conclusions

In this study, we explored the effects of Ti and Al concentrations on the formation of $L1_2$ nanoprecipitates and their influence on the mechanical properties of CoNiFeAlTi MPEAs using atomistic simulations. Our key findings are:

1- $L1_2$ nanoprecipitate formation: Using hybrid MD/MC simulations at 300 K and 1100 K, we observed the formation of Ni-rich $L1_2$ NPs in four MPEAs: $(CoNiFe)_{84}(Al_8Ti_8)$, $(CoNiFe)_{86}(Al_7Ti_7)$, $(CoNiFe)_{88}(Al_6Ti_6)$, and $(CoNiFe)_{94}(Al_4Ti_2)$ MPEAs. Among these, the $CoNiFe)_{86}(Al_7Ti_7)$ alloy demonstrated the highest concentration of $L1_2$ NPs, highlighting its potential for achieving a balanced combination of strength and ductility.

2- Enhanced tensile properties: The presence of $L1_2$ NPs significantly increased the yield strength and flow stress of the alloys. Specifically, the ordered phases containing $L1_2$ NPs in the $(CoNiFe)_{86}(Al_7Ti_7)$ alloy exhibited higher yield strength and flow stress compared to their random counterparts. The formation of more sessile Stair-rod dislocations in the alloy containing $L1_2$ may be responsible for the increased flow stress.

3- Dislocation-Nanoprecipitate interactions: Our analysis revealed that dislocations primarily shear through $L1_2$ NPs rather than bypass them via the Orowan mechanism. This dislocation-cutting mechanism, observed in NPs sized 4-8 nm, contributed significantly to the material's strengthening.

4- Stacking Fault Energy (SFE): Hybrid MD/MC results indicate that the formation of $L1_2$ NPs correlates with an overall increase in the calculated SFE of the alloy system. The substantial difference in SFE between the NPs and the FCC matrix enhanced dislocation pinning, leading to increased material strength.

In summary, our findings highlight the variation in the amount of $L1_2$ phase achieved by adjusting Al and Ti concentrations, identifying the optimal values for maximizing $L1_2$ phase formation. This study demonstrates a novel strategy for designing MPEAs with enhanced strength and ductility by leveraging low lattice mismatch and high stacking fault energy mismatch between precipitates and the matrix. The insights gained from this research lay a solid foundation for designing new MPEAs with tailored properties, effectively balancing strength and ductility through nanoscale precipitation engineering. This work not only advances the understanding of nanoprecipitate behavior in complex alloys but also provides practical guidance for future material design and optimization efforts

**Acknowledgments**

This research was funded by the European Union Horizon 2020 research and innovation program under NOMATEN Teaming grant (agreement no. 857470) and from the European Regional Development Fund via the Foundation for Polish Science International Research Agenda PLUS program grant No.MAB PLUS/2018/8. The publication was created within the framework of the project of the Minister of Science and Higher Education "Support for the activities of Centres of Excellence established in Poland under Horizon 2020″ under contract no. MEiN/2023/DIR/3795.

We acknowledge the computational resources provided by the High Performance Cluster at the National Centre for Nuclear Research in Poland.

**Data availability**

Data will be made available on request.
.